\documentclass[11pt,twoside]{article}
\usepackage{asp2004}
\usepackage{psfig}
\usepackage{epsf}
\usepackage{graphics}
\usepackage{lscape}
\def\ee #1   {\times 10^{#1}}          % \ee p       10^p
\def\ut #1 #2{ \, \mathrm{#1}^{#2}} % \ut unit p  unit^p
\def\u #1    { \, \mathrm{#1}}          % \u unit     unit
\def\msol    {\hbox{$M_\odot$}}

\def\edcomment#1{\iffalse\marginpar{\raggedright\sl#1\/}\else\relax\fi}
\reversemarginpar

\begin{document}
\title{Starburst Driven Thermal and Non-thermal Structures  
in the Galactic Center Region}
\author{F. Yusef-Zadeh(1), W. Cotton (2), J. Hewitt (1), C. Law (1), R. 
Maddalena (3) \& D. A. Roberts (1,4)}
%\author{W. Cotton (2)}
%\author{J. Hewitt (1)}
%\author{C. Law (1)}
%\author{R. Maddalena (2)} 
%\author{D. A. Roberts (1,3)}

\affil{(1) Dept.  Physics and Astronomy,
Northwestern University,  Evanston, IL.
60208}

\affil{(2) NRAO, 520 Edgemont Road, Charlottesville, VA 22903}

\affil{(3) NRAO, PO Box 2, Green Bank, WV 24944}

\affil{(4) Adler Planetarium \& Astronomy Museum,
1300 S. Lake Shore Drive, Chicago, IL. 60605}

\begin{abstract}

We briefly  review the prominent thermal and nonthermal sources near 
the Galactic center. These sources include  the young stellar clusters, 
the Sgr B complex  as well as the large-scale nonthermal filaments and 
lobes. 
Some of the recent radio images of this region based on VLA and Green Bank 
Telescope observations are also presented.  We then argue that the origin 
of the large-scale features within the inner two degrees of the Galactic 
center is tied to a past starburst activity in the nucleus of the Galaxy.
 
\end{abstract} \thispagestyle{plain}

\section{Sgr A*, Young Stellar Clusters, Nonthermal Filaments and the 
Galactic Center Lobe}

Like a jungle where many species evolve, share the same resources and 
interact with each other, the center of our Galaxy is occupied by an 
impressive collection of components  that coexist and interact with each 
other.  In this crowded environment, a clear understanding of the nature 
of these interacting components is not a trivial task. The difficulties 
within this environment such as crowding, a high visual extinction,  as 
well 
as unusual physical conditions 
are responsible for much debate on the physical nature and origin of many 
objects in this complex region. However, when a clear picture is 
revealed, 
the impact of such a study can be significant as it provides much insight 
into  the study of other regions and acts as a vehicle to explain 
the connection between  the physical 
processes operating in  the disk of the Galaxy and those within   
the nuclei of distant galaxies.

One of the components that has provided a great deal of excitment in 
recent years is the compact nonthermal source Sgr A$^*$ whose luminosity 
is thought to be due to accreting thermal winds from its neighboring 
cluster of massive stars. Sgr A* is considered to be a massive black hole 
at the dynamical center of the Galaxy. The revelation of a large 
concentration of dark matter 3-4 $\times 10^6$ \msol coincident with Sgr 
A* has created a much better picture of this object since its discovery 30 
years ago (Sch\"odel et al. 2003; Ghez et al. 2003). Presently, one of the 
key questions is whether the low luminosity of Sgr A$^*$  is due to the 
low 
accretion rate from mass-losing stars or the unusually high 
radiative deficiency of the accreting flow.

Another component  that has attracted much attention recently is the 
discovery of several clusters of young, massive stars within the projected 
distance of 30 pc from the Galactic center.  These objects are not 
run-of-the-mill clusters in the Galaxy and it is remarkable that at least 
three of such young and compact systems are found in a small volume of the 
Galaxy.  These high density stellar systems,  known as the Arches 
(G0.12+0.02), the 
Quintuplet (G0.15-0.05) and the central Sgr A clusters (e.g., IRS 16 and 
IRS 13) ,  consist of mainly 
O and WR stars 
with individual stellar masses greater than 20 \msol.
 The Arches cluster  contains more than 150 young,  
massive  stars with emission lines, all of which   are distributed within 
a few arcseconds 
of the 
core of the cluster (Figer et al. 1999). Recent theoretical studies 
of these types of massive clusters suggest that their short time scale 
for dynamical 
evolution  can lead to  the formation of intermediate  
mass  black holes via runaway merging (McMillan et al. 2004). In
fact,  a recent claim suggests  dynamical evidence for an intermediate 
mass black hole within  
the IRS 13 cluster near Sgr A$^*$ (Maillard et al. 2004).

These clusters provide an excellent laboratory to study both thermal and 
nonthermal processes. Multi-wavelength observations indicate ionized 
stellar winds arising from mass-losing WN and/or Of stars with mass-loss 
rates ranging between $10^{-5}$ to $10^{-4}$ \msol yr$^{-1}$ (Lang et al. 
2001a,b).  Most of the observations in radio, X-rays and near-IR 
wavelengths have concentrated on the study of thermal emission from ionized 
stellar winds in the cluster, however, WR+OB binary systems within these  
stellar clusters are also 
excellent sites for acceleration of particles to relativistic energies.
Theoretically, at the contact discontinuity where the winds of a binary 
system collide with each other, particles are accelerated by first order 
Fermi acceleration in shocks which then results in significant radiation 
(Dougherty et al. 2003). The spectrum of radio emission from binary stars 
could vary 
between fully thermal and nonthermal and many isolated WR-OB 
binary systems have displayed this characteristic. Although free-free 
absorption in these sources can be important, 
the relativistic particles can leave the binary systems and the clusters 
to be injected into the ISM.  Interestingly,
nonthermal emission has recently been detected from the 
Arches cluster. This implies that
 this cluster consists of a number of colliding wind 
binaries (Yusef-Zadeh et al. 2003). The additional role of nonthermal 
particles 
can be the production of nonthermal $\gamma$-ray and X-ray emission from 
upscattering of the intense radiation field as well as the irradiation of 
adjacent molecular clouds. In fact, the Arches cluster is surrounded by 
the 
fluorescent 6.4 keV line emission and also displaced  only by 
$\approx200''$ from the nominal position of the unidentified EGRET source 
3EG J1746--2851. The location and the spectrum of 
the $\gamma$-ray source is well within the 95\% error radius of 
$0.13^{\circ}$ 
and 
is 
consistent with  inverse Compton scattering.

On a large scale, two other components that have been recognized for more 
than 20 years are the striking nonthermal filaments and the puzzling 
 `` Galactic Center Lobe''. The filaments are found only 
within the inner 2$^0$ of the Galactic center and have  transverse 
dimensions 
that are roughly a fraction of a pc whereas their length is on  the order 
of tens of parsecs. The Galactic Center Lobe consists of two ``columns" of 
continuum emission with a degree scale ($\sim$150 pc) rising in the 
direction away from the Galactic plane. Within the region where both the 
Lobe and the nonthermal filaments are found, there is considerable 
amount of thermal ionized and dust emission associated with star forming 
regions.

In order to get a better understanding of the Lobe and the filaments and 
their relationship to star forming regions, we recently carried out 
multi-configuration VLA and multi-wavelength GBT observations of several 
degrees of the Galactic center. We used the wide field imaging technique 
to correct non-coplanar effects at 1.4 GHz based on 40 overlapping VLA 
pointings. Figure 1 shows a segment of the survey image based on combining 
the VLA and GBT data at 1.4 GHz. Figure 2 shows a segment of GBT data 
displaying  the Galactic Center Lobe at 5 GHz.

This study had led to several lines of evidence suggesting that the inner 
few hundred pcs of the Galaxy went through a burst of star forming 
activity  less than 10 million years ago.  One line of evidence comes 
from 
the number of young star clusters with similar ages of a few million years 
distributed in this region. The hallmark of an intense episode of  
star formation  
are  the luminous, compact, young star clusters.  Star cluster formation 
has also been considered an
 important mode of star formation in a high 
pressure environment in starburst galaxies. 
Numerical simulations of the 
evolution of massive 
star clusters within $\sim$200 pc of the Galactic center predict that the 
inner 200 pc of the Galaxy could harbor some 10 to 50 young star clusters 
similar to the Arches and the Quintuplet clusters 
(Portegies Zwart et al. 2002). However, the high 
visual extinction and source confusion will make it difficult to unravel 
the hidden star 
clusters in this region (Law \&  Yusef-Zadeh 2004). An example in 
which  star 
clusters may be hidden in a highly obscured region is 
the Sgr B  complex. This complex consists  of an evolved extended HII 
region Sgr B1 and  the young Sgr B2  source whose  emission is 
dominated by  compact,  bright continuum HII regions. Sgr B2 may 
signify the most spectacular on-going 
star forming region in the Galaxy containing more than 50 compact HII 
regions, many of which are excited by young massive stars.
 In the context 
of the starburst model in the Galactic center region, the Sgr B region  is 
an  example of a starburst
that took place  few million years ago. However,  because of its  
dense and 
massive  
molecular environment,  
induced star formation  has continued until 
present by expanding HII regions.  Figure 3 shows 
a 1.4 GHz continuum  image of the Sgr B2 complex.

Our 20cm survey has found a large fraction of the filaments showing  
jet-like morphology   and 
a wide range of orientations with respect to the Galactic plane. Figure 4 
shows a schematic diagram of the distribution of the filaments where more 
than 80 filaments are drawn. The longest filaments run 
roughly perpendicular to the Galactic plane whereas the short filaments do 
not show a preferred orientation. 
It is not a coincidence  that 
the largest concentration of the filaments are populated in star forming 
regions and are found only  within the inner two degrees of the 
Galactic center.   These observations suggest 
the origin of the magnetic fields tracing the filaments is likely to be 
consistent with being local rather than global (LaRosa et al. 2004; 
Yusef-Zadeh 2003). 
Specifically, the filaments may be the result of nonthermal particles
generated in the colliding winds of WR and OB stars.
In addition, 
the same WR-OB binary systems can be responsible for much of dust 
formation found in the Lobe. Another speculation   for the origin of the 
filaments is black holes,  formed within massive young 
clusters as a result of runaway merging,  are responsible for  launching  
narrow filaments. 

Other independent studies of the ISM of the Galactic center region 
support a similar picture of starburst activity (Bland-Hawthorn \& Cohen 
2003; Rodriguez-Fernandez \& Martin-Pintado 2004). 
In particular, recent ISO observations of ionized gas in this region 
show excitation and ionization parameters that are similar to some 
low-excitation starburst galaxies. 
We believe the unusual 
collection of remarkable thermal and nonthermal components found in the 
Galactic center region can be viewed as a manifestation of windy massive stars 
affecting their surrounding ISM in a starburst episode. 
In particular, a WR-type phenomenon may be a thread that connects
 the accreting material onto Sgr A$^*$, 
young stellar clusters, the nonthermal filaments and the Galactic Center 
Lobe.  
Future study of 
this region can shed light on our understanding of more energetic a 
WR-type 
phenomena in distant galaxies and conversely, studying
distant galaxies can aid our understanding of energetic WR-phenomena.

\section{References}
\begin{quote}
Bland-Hawthorn, J. \& Cohen, M. 2003, ApJ, 582, 246\\
Dougherty, S.M. et al. 2003, A\&A, 217-233\\
Figer et al. 1999, ApJ, 514, 202\\
Ghez, A. et al. 2003 ApJ, 586, L127\\
Lang, C.C., Goss, W.M. \& Morris, M.  2001a, AJ, 121, 2681\\
Lang, C.C., Goss, W.M. \& Rodriguez, L.F.  2001b, ApJ, 551, L143\\
LaRosa, T. N.,  Nord, M. E., Lazio, J.T. \&  Kassim, N.E.
2004, ApJ, 607, 302\\
Maillard, J.P.  2004, A\&A, 423, 155\\
McMillan, S. et al. 2004, in "Formation and Evolution of Massive Young 
Star Clusters," eds. H.J.G.L.M. Lamers, A. Nota \& L.J. Smith, in press 
(astro-ph0411166)\\
Law, C.  and Yusef-Zadeh, F. 2004, ApJ, 611, 858\\
Portegies Zwart, S.F., Makino, J., McMillan, L.W., Hut, P.
2002, ApJ,  565,  265\\
Rodriguez-Fernandez, N.J., Martin-Pintando, J. 2004, A\&A, in press 
(astro-ph/0409334)\\
Sch\"odel, R., Ott, T., Genzel, R. et al. 2003, ApJ, 596, 1015\\
Yusef-Zadeh, F. 2003, ApJ, 598, 325\\
Yusef-Zadeh, F. 2003, ApJ, 598, 325\\
Yusef-Zadeh, F., Hewitt, J. \& Cotton, W. 2004, ApJS, in press\\
\end{quote}

\vfill\eject

\begin{quote}
\begin{figure}[!ht]
%\plotone{fig1.cps}
% \plotfiddle{PSFILE}{VSIZE}{ROTANG}{HSCALE}{VSCALE}{HTRANS}{VTRANS}
%\plotfiddle{fig1.cps}{8in}{-90}{80}{80}{-300}{600}
\caption{A combined 20cm image from the VLA and 
GBT with a resolution of 30$''$. The features 
near the top are some of the nonthermal
radio filaments. Some of the known 
 supernova remnants (SNRs) and
HII regions are also shown. }
\end{figure}
\end{quote}

\begin{quote}
\begin{figure}[!ht]
%\plotone{fig2.cps}
\caption{A 6cm image of the Galactic center region 
based on GBT observation with a resolution of 3$'$. The 
outline of the Galactic 
Center Lobe is  drawn. The Galactic plane runs horizontally.} 
\end{figure}
\end{quote}

\begin{quote}
\begin{figure}[!ht]
%\plotone{fig3.cps}
%\plotfiddle{fig3.cps}{8in}{-90}{80}{80}{-300}{600}
\caption{A 20cm image of the Sgr B complex region 
 with a resolution of 2.5$''\times1.7''$. Sgr B2 and Sgr B1 lie to the 
northeast and southwest, respectively. The 
Galactic plane 
run diagonally.
} \end{figure}
\end{quote}

\begin{quote}
\begin{figure}[!ht]
%\plotone{fig4.ps}
\caption{A schematic diagram of all the identified radio filaments. 
The position of Sgr A$^*$ is presented by a star.
The gray background circles show the extent of the
surveyed region.
}
\end{figure}
\end{quote}
                                                                             
\end{document}